\documentstyle[sprocl,graphicx,amssymb]{article}
\DeclareGraphicsExtensions{.eps}

\newcommand{\eprint}[2]{{hep-#1/}#2.}

\newcommand{\jrnl}[4]{{#1} {\bf #2} (#3) #4.}
\newcommand{\jrnlk}[4]{{#1} {\bf #2} (#3) #4,}
\newcommand{\jrnls}[4]{{#1} {\bf #2} (#3) #4;\\}
\newcommand{\EPJC}{{\em Eur.\ Phys.\ J.\ }{\bf C}}

\newcommand{\JHEP}{\em JHEP}

\newcommand{\NPB}{{\em Nucl.~Phys.\ }{\bf B}}

\newcommand{\PLB}{{\em Phys.~Lett.\ }{\bf B}}

\newcommand{\mean}[1]{\left< #1 \right>}
\newcommand{\an}{\overline{\alpha}_0}

\newcommand{\as}{\alpha_s}
\newcommand{\asmz}{\alpha_s(M_Z)}
\newcommand{\ee}{\mbox{$e^+e^-$~}}
\newcommand{\fmean}{\mean{F}}
\newcommand{\fpert}{\fmean^{\rm pert}}
\newcommand{\fpow}{\fmean^{\rm pow}}
\newcommand{\gev}{\,{\rm GeV}}

\newcommand{\hftwo}{\hspace*{\fill}}

\newcommand{\mi}{\mu_{\scriptscriptstyle I}}

\begin{document}

\title{EVENT SHAPES AND POWER CORRECTIONS\\
  IN \boldmath$ep$\unboldmath\ DIS AT HERA}

\author{K.~RABBERTZ} \address{I.~Physikalisches Institut, RWTH Aachen,
  D-52056 Aachen, Germany} \author{U.~WOLLMER} \address{DESY Hamburg,
  D-22603 Hamburg, Germany\\[2ex](on behalf of the H1 and ZEUS collaborations)}


\maketitle\abstracts{%
  \noindent
  Deep-inelastic $ep$ scattering data, taken with the H1 and ZEUS
  detectors at HERA, are used to study the means and distributions of
  the event shape variables thrust, jet broadening, jet mass,
  $C$-parameter and two kinds of differential two-jet rate.  The data
  cover a range of the four-momentum transfer $Q$, taken to be the
  relevant energy scale, between $7\gev$ and $141\gev$.  The $Q$
  dependences are compared with second-order calculations of
  perturbative QCD\@.  Power law corrections are applied to account
  for hadronization effects.}

\section{Introduction}

Event shapes are observables designed to study the influence of the
strong interaction on hadronic final states by characterizing
deviations from the pencil-like structure to be expected within the
framework of the quark parton model.  Measurements have shown,
however, that even at energies as high as the $Z$ mass
non-perturbative effects have to be taken into account.  The search
for a better theoretical understanding of these non-perturbative
contributions has prompted a revival of interest in event
shapes.\cite{pc}

\section{The Measured Data}

A suitable frame of reference to study event shapes in DIS is the
Breit frame which maximizes the separation between the current jet
from the struck quark and the proton remnant. In this frame, the
exchanged gauge boson is purely space-like with four-momentum $q =
\{0,0,0,-Q\}$. Within the framework of the quark parton model, the
incoming quark with longitudinal momentum $p^{\rm in}_{q\,z} = Q/2$ is
back-scattered into the current hemisphere ($z<0$) with $p^{\rm
  out}_{q\,z} = -Q/2$ while the proton fragments into the opposite
direction (remnant hemisphere, $z>0$).  The studies presented by the
H1 and ZEUS collaborations~\cite{H1EPJC,ZEUSMoriond} investigate five
event shapes which are confined to the current region: thrust with
respect to the thrust axis ($\tau_C$, $\tau_m$), thrust and jet
broadening with respect to the boson axis ($\tau$, $\tau_z$; $B$,
$B_c$), jet mass ($\rho$) and the $C$-parameter~($C$).  In order to keep
these variables infrared-safe, the total energy in the current
hemisphere has to exceed 20\% (H1) or 6\% (ZEUS) of the value expected
from the quark parton model ($Q/2$).  In addition, H1 has studied two
kinds of differential two-jet rates, i.e.\ the transition value from
$(2+1)$ to $(1+1)$ jets,\footnote{The $+1$ denotes the proton remnant
  jet.}  exploiting the full phase space: the factorizable JADE
($y_{fJ}$) and the $k_t$ algorithm ($y_{k_t}$).

\begin{figure}[tb]
  \hftwo\includegraphics[width=0.75\textwidth]{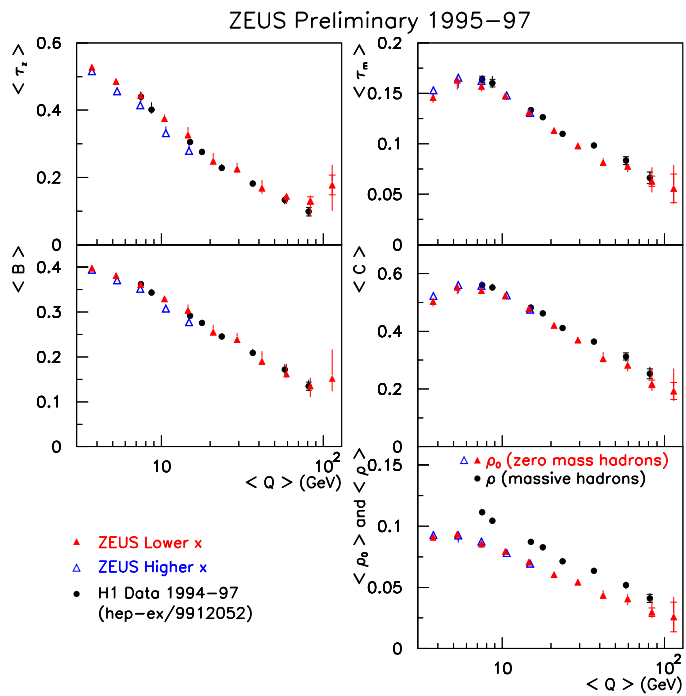}\hftwo
  \caption{Comparison of unfolded event shape means as measured by
    H1 and ZEUS.}
  \label{means}
\end{figure}

The analyzed data cover a kinematic range in $Q$ from $7\gev$
($9\gev$) up to $100\gev$ ($141\gev$) for H1 (ZEUS) and are integrated
over the scaling variable Bj{\o}rken $x$ in H1 whereas ZEUS employs a
binning in $x$ as well.  Note that while unfolding the influence of
the detector on the measured data ZEUS additionally corrects as
proposed~\cite{GSpriv} for mass effects by assuming all hadrons to be
massless (P scheme).  This leads to differences for the means of the jet mass
between ZEUS ($\rho_0$) and H1 ($\rho$) as shown in fig.~\ref{means}.
Thrust, jet broadening and the $C$-parameter are unaffected because they
are derived from three-momenta and are in good agreement between the
two collaborations.

\section{Power Corrections to Mean Values}

Hadronization effects on an event shape, generically labelled as $F$,
are treated within the concept of power corrections which appear as
additive terms to the perturbative predictions for the mean values and
are proportional to $1/Q^p$ with exponents $p=1$ ($p=2$ for
$y_{k_t}$):
\begin{equation}
  \fmean = \fpert + \fpow = \fpert + a_F \cdot {\cal P}
  \label{meansum}
\end{equation}
with $a_F$ being a $F$ dependent calculable coefficient.  A simple
form of ${\cal P} \propto {\rm const}/Q^p$ is not sufficient to
describe the data~\cite{H1EPJC}. A more sophisticated
approach~\cite{pc} introduces for $p=1$
\begin{equation}
  {\cal P} = 1.61 \frac{\mi}{Q} \left[ \an(\mi) - \as(Q) -
    1.22 \left(\ln\frac{Q}{\mi} + 1.45 \right) \as^2(Q) \right]
  \label{pc}
\end{equation}
where the renormalization scale has been identified with $Q$ and
$\mi=2\gev$ is the infrared-matching scale below which the strong
interaction is parameterized by a non-perturbative parameter $\an$
that corresponds to an average effective strong coupling in the
infrared region.

\begin{figure}
  \hftwo\includegraphics[width=0.33\textwidth]{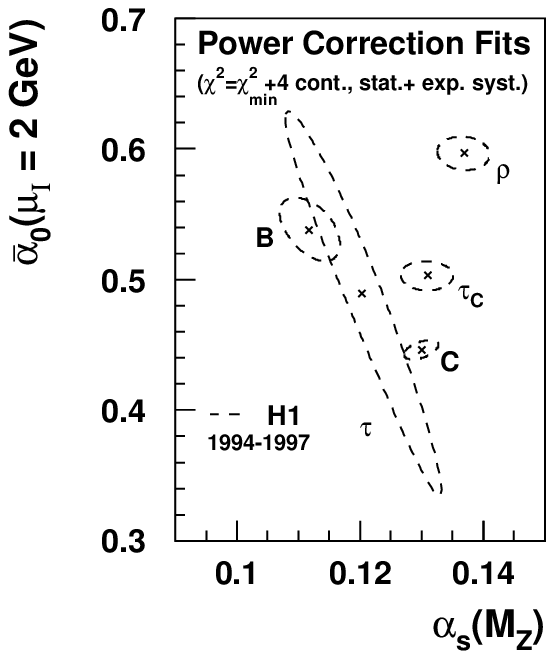}\hftwo%
  \includegraphics[width=0.33\textwidth]{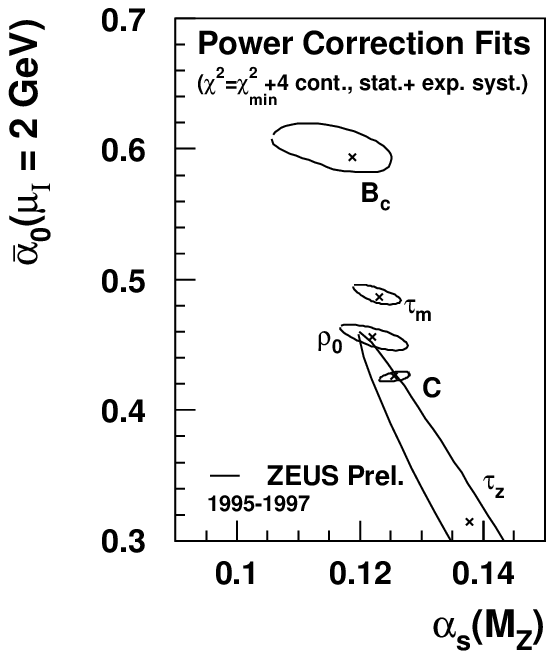}\hftwo%
  \includegraphics[width=0.33\textwidth]{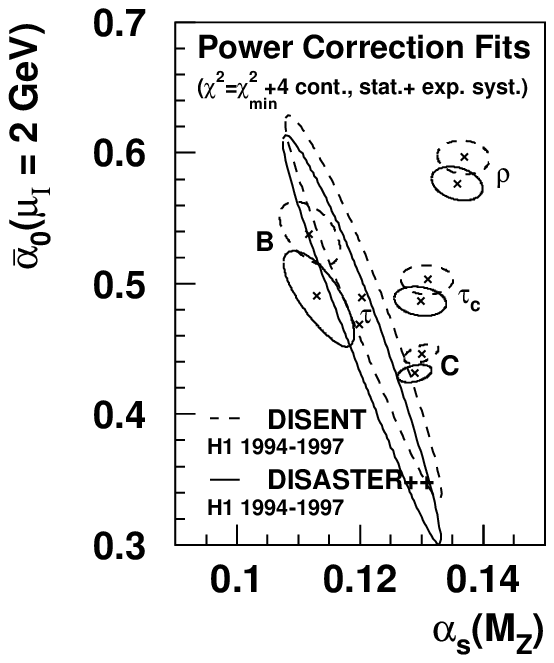}\hftwo
  \caption{Comparison of power correction fits for H1 (left) and ZEUS data
    (middle), differences between the fits for H1 data using either DISENT or
    DISASTER++ (right).}
  \label{ellipses}
\end{figure}

The results of fits of $\an$ and $\asmz$ according to
eqs.~\ref{meansum} and~\ref{pc} are summarized in fig.~\ref{ellipses}
in the form of contours of $\chi^2(\as,\an)=\chi_{\rm min}^2 + 4$.
For $\tau_C$ ($\tau_m$) and $C$ there is fair agreement whereas due to
the different treatment of hadron masses in the correction procedure
the jet masses $\rho$ ($\rho_0$) differ.  Performing the fit
simultaneously for two different regions in $x$ as done by ZEUS is
more problematic and leads to the larger discrepancies which are
observed for $\tau$ and $B$.  For a comparison to results from \ee
annihilation see~\cite{biebel}.

On the right hand side of fig.~\ref{ellipses} the influence of
discrepancies~\cite{dasgupta} that have been observed for the two pQCD
programs DISENT~\cite{disent} and DISASTER++~\cite{disaster} is
demonstrated.  DISASTER++ leads in general to a somewhat smaller $\an$
and to slightly more consistent values of $\asmz$.

Similar fits have been performed by H1~\cite{H1EPJC} for the
differential two-jet rates $y_{fJ}$ and $y_{k_t}$ both of which
exhibit much smaller hadronization corrections than the other event
shapes. In the case of $y_{fJ}$ the suggested value of $a_{y_{fJ}}=1$,
i.e.\ the same as for thrust, can be excluded from the data. For the
$k_t$ algorithm the coefficient of the $p=2$ power correction is not
known and due to large correlations for a three parameter fit it can
only be stated that the data are consistent with quadratic power law
corrections.

\section{Power Corrections to Distributions}

\begin{figure}
  \hftwo\includegraphics[height=6cm]{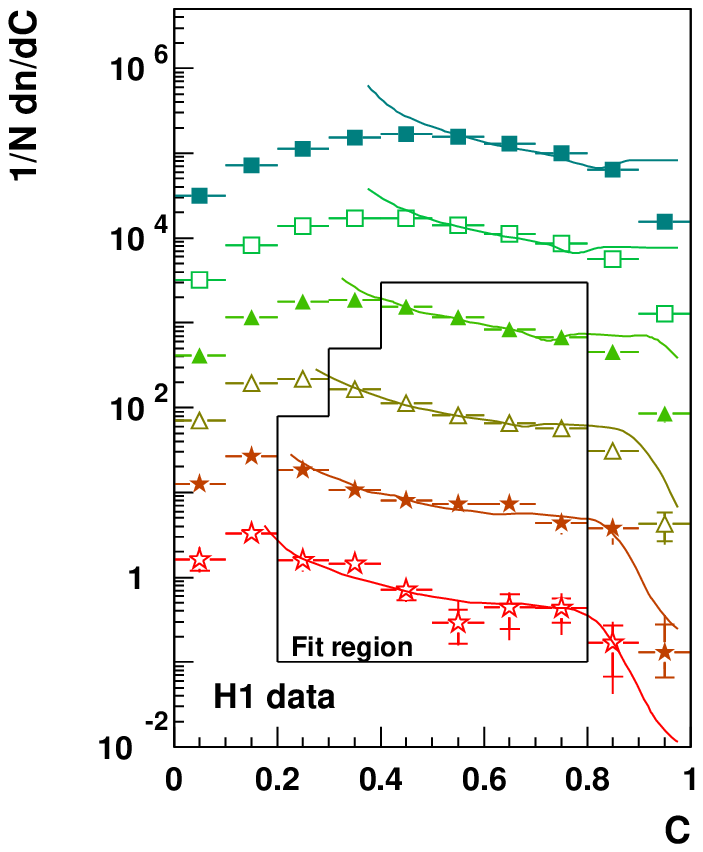}\hftwo%
  \includegraphics[height=6cm]{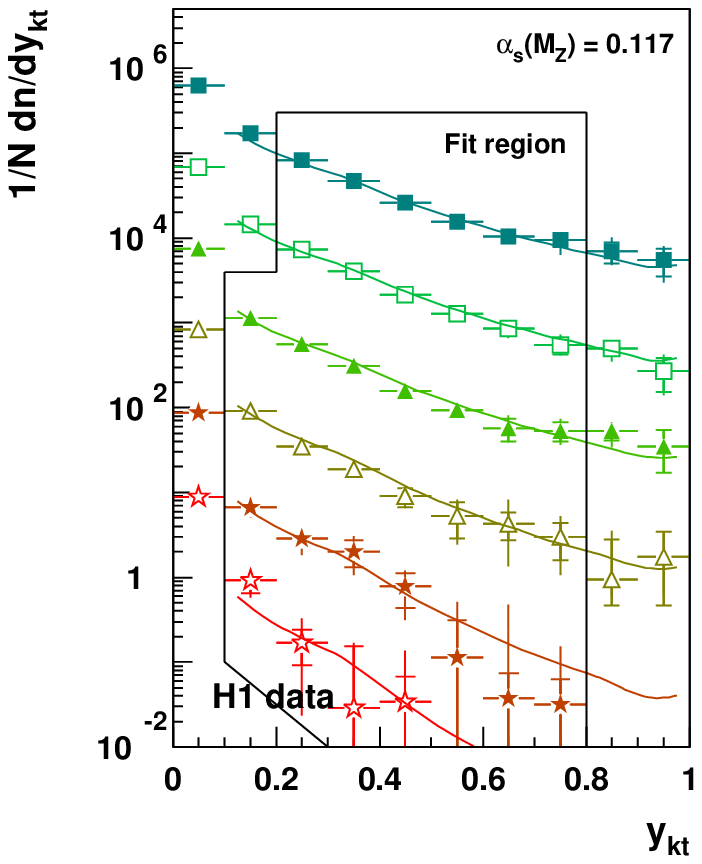}\hftwo
  \caption{Fits of differential distributions for $C$ with power correction
    (left) and for $y_{k_t}$ without (right). The different symbols
    label the binning in $Q$ with $\mean{Q}$ ranging from $15.0\gev$
    ($\blacksquare$) up to $81.3\gev$ ($\star$).}
  \label{distributions}
\end{figure}

If distributions in the event shapes confined to the current
hemisphere are studied the power corrections can (except for $B$) be
written as
\begin{equation}
  \frac{1}{\sigma_{\rm tot}}\,\frac{d\sigma(F)}{dF} =
  \frac{1}{\sigma_{\rm tot}}\,\frac{d\sigma^{\rm pert}(F-a_F{\cal P})}{dF}
  \label{distshift}
\end{equation}
provided $\mi/Q \ll F$. The power term ${\cal P}$ is exactly the
same as that of
eq.~\ref{pc}.  Fig.~\ref{distributions} shows fits of the
next-to-leading order calculations for the distributions in $C$ and $y_{k_t}$,
with and without power correction, respectively. In general, these
exercises lead to inconsistent results between the outcome of fits to
the means and the distributions of the same observable, e.g.~$C$ gives
for ($\an$, $\asmz$) $(0.45,0.130)$ from $\mean{C}$ and $(0.62,0.131)$
from $d\sigma/dC$.  A scheme for matching resummed and next-to-leading
order distributions, which for $ep$ DIS has only recently become
available~\cite{dasgupta}, may be necessary. Only in case of the
differential two-jet rates, where hadronization corrections are smaller,
can reasonable fits at sufficiently high $Q$ be obtained while
neglecting power corrections completely. An example is given in
fig.~\ref{distributions} for $y_{k_t}$.

Including resummed predictions, the same techniques have been
successfully applied in \ee annihilation~\cite{biebel}.

\section{Conclusion}

In summary, the power correction approach to hadronization appears to
work well in $ep$ DIS and \ee annihilation, although some theoretical
and experimental questions remain. An approximately universal value of
$\an \approx 0.5\pm 20\%$ can be deduced. With further progress in
resummed predictions, a fruitful future lies ahead of us.


\section*{References}
\begin{thebibliography}{99}
\bibitem{pc}%
  Yu.L.~Dokshitzer, B.R.~Webber, \jrnlk{\PLB}{352}{1995}{451}
  Yu.L.\ Dokshitzer, G.~Marchesini, B.R.~Webber, \jrnl{\NPB}{469}{1996}{93}
\bibitem{H1EPJC}%
  H1 Coll., C.~Adloff et al., \jrnl{\EPJC}{14}{2000}{255}
\bibitem{ZEUSMoriond}%
  G.~McCance, in Proc.~{\em Rencontres de Moriond}, Les Arcs, France, 2000.
\bibitem{GSpriv}%
  G.P.~Salam, private communication. 
\bibitem{biebel}%
  O.~Biebel, these Proceedings.
\bibitem{dasgupta}%
  V.~Antonelli, M.~Dasgupta, G.P.~Salam, \jrnls{\JHEP}{02}{2000}{001}
  M.~Dasgupta, these Proceedings.
\bibitem{disent}%
  S.~Catani, M.H.~Seymour, \jrnl{\NPB}{485}{1997}{291}
\bibitem{disaster}%
  D.~Graudenz, \eprint{ph}{9710244}
\end{thebibliography}
\end{document}